\documentclass[a4paper,fleqn,usenatbib,useAMS]{mnras}
\usepackage{aas_macros}
\usepackage{graphicx}
\usepackage{psfig}
\usepackage{amsmath}	
\usepackage{amssymb}	

\newcommand{\ltsima} {$\; \buildrel < \over \sim \;$}  
\newcommand{\gtsima} {$\; \buildrel > \over \sim \;$}  
\newcommand{\lta} {\lower.5ex\hbox{\ltsima}}  
\newcommand{\gta} {\lower.5ex\hbox{\gtsima}}  

\newcommand{\ergs}{\>{\rm erg}\,{\rm s}^{-1}}

\newcommand{\loiii}{L$_{\rm{\tiny{ [O~III]}}}$}

\title[VLA observations of FR\,0 radio galaxies]{High-resolution VLA
  observations of FR\,0 radio galaxies: \\ properties and nature of compact
  radio sources.} 
\author[R.~D. Baldi, A. Capetti and G. Giovannini] {Ranieri
  D. Baldi$^{1}$\thanks{E-mail: r.baldi@soton.ac.uk}, Alessandro
  Capetti$^{2}$, Gabriele Giovannini$^{3,4}$ 
\\ $^{1}$ School of Physics and Astronomy, University of Southampton,
Southampton, SO17 1BJ, UK\\ 
$^{2}$ INAF - Osservatorio Astrofisico di
Torino, Strada Osservatorio 20, I-10025 Pino Torinese, Italy\\ 
$^{3}$ Dipartimento di Fisica e Astronomia, Universit\`a di Bologna,  via Gobetti 93/2, 40129 Bologna, Italy\\
$^{4}$ INAF - Istituto di Radio Astronomia, via P. Gobetti 101, I-40129, Bologna, Italy}
\begin{document}


\pagerange{\pageref{firstpage}--\pageref{lastpage}} \pubyear{2014}

\maketitle

\label{firstpage}

\begin{abstract}

We present the results of Karl G. Jansky Very Large Array (VLA) observations
to study the properties of FR~0 radio galaxies, the compact radio sources
associated with early-type galaxies which represent the bulk of the local
radio-loud AGN population. We obtained A-array observations at 1.5, 4.5, and
7.5 GHz for 18 FR~0s from the FR0{\sl{CAT}} sample: these are sources at
$z<0.05$, unresolved in the FIRST images and spectroscopically classified as
low excitation galaxies (LEG). Although we reach an angular resolution of
$\sim$0.3 arcsec, the majority of the 18 FR~0s is still unresolved. Only four
objects show extended emission. Six have steep radio spectra, 11 are flat
cores, while one shows an inverted spectrum. We find that 1) the ratio between
core and total emission in FR~0s is $\sim$30 times higher than in FR~I and 2)
FR~0s share the same properties with FR~Is from the nuclear and host point of
view. FR~0s differ from FR~I only for the paucity of extended radio
emission. Different scenarios were investigated: 1) the possibility that all
FR~0s are young sources eventually evolving into extended sources is ruled out
by the distribution of radio sizes; 2) similarly, a time-dependent scenario,
where a variation of accretion or jet launching prevents the formation of
large-scales radio structures, appears to be rather implausible due to the
large abundance of sub-kpc objects 3) a scenario in which FR~0s are produced
by mildly relativistic jets is consistent with the data but requires
observations of a larger sample to be properly tested.

\end{abstract}

\begin{keywords}
galaxies: active $-$ galaxies: elliptical and lenticular, cD $-$ galaxies: nuclei - galaxies: jets $-$ radio continuum: galaxies
\end{keywords}

\section{Introduction}
\label{intro}

Among the variety of observed morphologies of radio-emitting Active
Galactic Nuclei (AGN) in the local Universe, the most common one is
the presence of a single compact emitting region
\citep{baldi10a}. This conclusion could be derived only after the
advent of deep large area radio surveys in opposition to high-flux
limited sample studies (such as the 3C, 2Jy, and B2 catalogues
\citealt{bennett62a,wall85,colla75}) which typically selected radio
sources extending on scale of many kpc and belonging to the Fanaroff
\& Riley classes I and II. The cross-match of optical and radio
surveys (SDSS, NVSS, and FIRST dataset, SDSS/NVSS sample,
\citealt{best05a,best12}) showed that compact radio sources, at
5$\arcsec$ resolution, represent the vast majority of the local radio
AGN population
\citep{baldi09,baldi10a,sadler14,banfield15,whittam16,whittam17,miraghaei17,lukic18}. Earlier
radio studies
\citep{rogstad69,heeschen70,ekers74,wrobel91b,slee94,giroletti05}
already pointed out that most of the radio sources in the local
universe are flat-spectrum compact sources, but the attention of the
community has been mainly focused on the study of extended radio
sources (FR~I/FR~II).

Compact radio sources can potentially be associated with different
classes of AGN, including radio-quiet AGN, compact steep-spectrum
sources (CSS) and blazars. By using multiwavelength data provided by
SDSS, \citet{baldi10a} selected from the SDSS/NVSS sample
\citep{best05a,best12} a large population of massive red early-type
galaxies (ETGs) associated with compact radio sources. These objects
have been named {\it FR0s} by \citet{ghisellini11} as a convenient way
of linking the compact radio sources seen in nearby galaxies into the
canonical Fanaroff-Riley classification scheme
\citep{sadler14}. Adopting these selection criteria, FR~0s form a
rather homogeneous population of low-luminosity radio galaxies
\citep{baldi18a}. FR~0s appear indistinguishable from low-power
FR~I/FR~II LEG radio sources \citep{capetti17a,capetti17b}, sharing
similar range of AGN bolometric luminosities, host galaxy properties
and BH masses: apparently, the only feature which characterise FR~0s
from the other FR classes is their lack of substantial extended radio
emission.

Since this vast population is still virtually unexplored, \citet{baldi15}
carried out a pilot study of Karl G. Jansky Very Large Array (VLA) radio
imaging of a small sample of FR~0s. The main result is that most of the
sources still appears compact at higher resolution, with $\sim$80 per cent of
the total radio emission unresolved in the core (i.e., they are highly
core-dominated). The few extended sources show a symmetric radio
structure. Their radio spectra generally are flat or steep, but with an
emerging flat core at higher radio frequencies. In addition, these sources
show radio core energetics, line and X-ray luminosities \citep{torresi18},
similar to FR~Is. \citet{baldi15} conclude that FR~0s are able to launch a jet
whose unresolved radio core base appears indistinguishable from those of
FR~Is, but not emitting prominently at large scales. This radio behaviour is
similar to what is observed in nearby giant ETGs which harbour low-power RL
AGN (10$^{36-38}$ erg s$^{-1}$; \citealt{baldi09,baldi16}), named Core
Galaxies (CoreG, \citealt{balmaverde06core}) or very low power (10$^{33-38}$
erg s$^{-1}$) LINERs recently studied with eMERLIN \citep{baldi18b} and VLA
\citep{nyland16}.  These sources exhibit compact radio emission at the VLA
resolution, only occasionally associated with diffuse extended emission
\citep{filho00,filho02,falcke00,nagar00,nagar02}, and consistent with a FR~0
classification.

The multiband properties of FR~0s indicate that their compactness and
high core dominance are genuine and not due to a geometric effect
\citep{baldi15,torresi18}. All these characteristics point out the
uniqueness of the radio properties of the FR~0s as a stand-alone
class, different from the other FR classes, blazars, CSS and
radio-quiet AGN.

We proposed various explanations to account for the radio properties
of FR~0s. Their small size might indicate the youth of their radio
activity, but \citet{baldi18a} showed that a scenario in which all
FR~0s eventually evolve into extended radio sources, cannot account
for the space numbers of different FR classes. However, other open
possibilities are: i) FR0s could be short-lived and/or recurrent
episodes of AGN activity, not long enough for radio jets to develop at
large scales \citep{sadler14}, ii) FR0s produce slow jets, possibly
due to small BH spin, experiencing instabilities and entrainment in
the dense interstellar medium of the host galaxy corona that causes
their premature disruption \citep{baldi09,bodo13}, and iii) the
differences between FR~0 and extended radio galaxies are due to a
distinct environment.

In order to explore the FR~0 with a more systematic approach we selected a
homogeneous and well defined sample of such sources, named FR0{\sl{CAT}}
\citep{baldi18a}. The FR0{\sl{CAT}} includes 104 compact radio sources
selected by combining observations from the NVSS, FIRST, and SDSS surveys (see
Appendix A for an update of the catalogue). We included in the catalogue the
sources with redshift $\leq 0.05$ and with an optical spectrum characteristic
of low excitation galaxies. We imposed a limit on their deconvolved angular
sizes of 4$\arcsec$, corresponding to a linear size $\lesssim$ 5 kpc, based on
the FIRST images. Their FIRST radio luminosities at 1.4 GHz are mostly in the
range $10^{38} \lesssim \nu L_{1.4} \lesssim 10^{40} \ergs$. The FR0{\sl{CAT}}
hosts are mostly (86 \%) luminous ($-21 \gtrsim M_r \gtrsim -23$) red early
type galaxies with black hole (BH) masses $10^8 \lesssim M_{\rm BH} \lesssim
10^9 M_\odot$ with a small tail down to $10^{7.5} M_\odot$.

Compactness is not a well defined property as it depends on the resolution,
sensitivity and the frequency of the available observations. Furthermore, we
lack of any information on their radio spectral shape. For these reasons we
started a comprehensive study of FR0{\sl{CAT}} sources with VLA observations
at higher frequency and resolution with respect to FIRST, in order to explore
their morphology at different scales and wavelengths.  Following the pilot
study \citep{baldi15}, here we present VLA A-array radio observations for 18
sources at 1.5, 4.5 and 7.5 GHz reaching a resolution of $\sim$0.3 arcsec to:
i) study the extended emission, morphology and asymmetry ; ii) resolve the
radio cores; iii) derive the radio spectral distributions.

The paper is organised as follows. In Sect. \ref{sample} we define the sample
and present the new VLA observations for 18 sources.  In Sect~\ref{results}
we analyse the radio and spectro-photometric properties of the sample which
are discussed in Sect.~\ref{discussion}. The summary and conclusions to our
findings are given in Sect.~\ref{conclusion}. We also present a revision of
the FR0{\sl{CAT}} sample in Appendix A.

\begin{figure}
\includegraphics[scale=0.55]{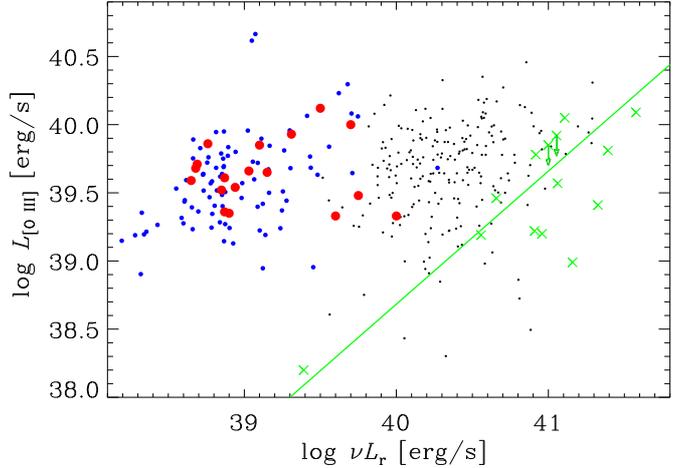}
\caption{Total radio luminosity at 1.4 GHz from NVSS vs [O~III] line
  luminosity (erg s$^{-1}$).  The large red dots represent the sub-sample of
  18 galaxies presented in this paper, the medium blue points represent the
  104 sources forming FR0{\sl{CAT}}. The small black dots are the FR~I radio
  galaxies of the FRI{\sl{CAT}} \citep{capetti17a} while the green crosses are
  the FRI in the 3C sample. The solid line represents the correlation between
  line and radio-luminosity derived for the 3CR/FR~I sample
  \citep{buttiglione10}.}
\label{lrlo3}
\end{figure}

\begin{table*}
\caption{The sample}
\begin{tabular}{|l|l|l|c|c|c|c|r}
\hline
  SDSS               &  name          & \,z      &   L$_{\rm [O~III]}$  & M$_{\rm BH}$
  & L$_{\rm NVSS}$        & L$_{\rm FIRST}$ & L$_{\rm core}$   \\
\hline
J090734.91+325722.9  &                &   0.049  &   39.33 &   7.7 &  39.60 &  39.52   &  $<$39.40   \\  
J093003.56+341325.3  & MCG~+06-21-042 &   0.042  &   39.93 &   8.4 &  39.31 &  39.24   &  $<$39.59   \\   
J093938.62+385358.6  &                &   0.046  &   39.59 &   8.1 &  38.65 &  38.62   &  39.22      \\ 
J094319.15+361452.1  & NGC~2965       &   0.022  &   39.85 &   7.9 &  39.10 &  39.05   &  $<$40.36   \\  
J101329.65+075415.6  &                &   0.046  &   39.86 &   8.7 &  38.76 &  38.68   &  39.18      \\  
J102544.22+102230.4  &                &   0.045  &   39.48 &   8.9 &  39.75 &  39.69   &  40.62      \\  
J104028.37+091057.1  & NGC~3332       &   0.019  &   39.54 &   8.3 &  38.94 &  38.86   &  $<$39.19   \\ 
J113637.14+510008.5  & MCG~+09-19-133 &   0.050  &   39.35 &   8.3 &  38.90 &  38.80   &  $<$38.71   \\  
J121329.27+504429.4  & NGC~4187       &   0.031  &   40.12 &   8.7 &  39.50 &  39.49   &  40.09      \\   
J123011.85+470022.7  & MCG~+08-23-044 &   0.039  &   40.00 &   8.4 &  39.70 &  39.63   &  40.13      \\   
J150808.25+265457.6  &                &   0.033  &   39.36 &   8.1 &  38.87 &  38.85   &  39.38      \\ 
J152010.94+254319.3  & MCG~+04-36-038 &   0.034  &   39.52 &   8.6 &  38.85 &  38.79   &  39.43      \\  
J153016.15+270551.0  &                &   0.033  &   39.71 &   8.2 &  38.69 &  38.66   &  39.93      \\   
J155951.61+255626.3  & IC~4587        &   0.045  &   39.33 &   8.5 &  40.00 &  39.28   &  39.68      \\  
J162146.06+254914.4  &                &   0.048  &   39.61 &   8.1 &  38.87 &  38.79   &  $<$38.96   \\  
J162846.13+252940.9  &                &   0.040  &   39.65 &   8.5 &  39.15 &  39.14   &  39.87      \\  
J165830.05+252324.9  &                &   0.033  &   39.68 &   8.3 &  38.68 &  38.71   &  39.41      \\  
J170358.49+241039.5  & UGC~10678      &   0.031  &   39.66 &   8.8 &  39.03 &  38.84   &  $<$38.85   \\   
  \hline
\end{tabular}
\label{tab1}

\smallskip
\small{Column description: (1) source SDSS name; (2) other name; (3)
  redshift; (4) logarithm of the [O~III] line luminosity (erg
  s$^{-1}$); (5) logarithm of the BH mass (M$_{\odot}$); (6 and 7)
  logarithm of the radio luminosity from NVSS and FIRST (erg s$^{-1}$); (8) logarithm
  of the radio core luminosity (erg s$^{-1}$).}
\end{table*}

\begin{table}
  \caption{Properties of of the JVLA images}
\begin{tabular}{|l|c|c|c|c|c|}
  \hline
  ID  & beam  &  rms &  ID  & beam  &  rms \\

  \hline
J0907+32     & 2.1$\times$1.4 & 0.041 & J1230+47     & 1.4$\times$1.3 & 0.025 \\
             & 0.7$\times$0.5 & 0.014 &              & 0.5$\times$0.5 & 0.020 \\
             & 0.4$\times$0.3 & 0.013 &              & 0.3$\times$0.3 & 0.035 \\
J0930+34     & 1.8$\times$1.3 & 0.062 & J1508+26     & 1.9$\times$1.4 & 0.028 \\
             & 0.7$\times$0.5 & 0.021 &              & 0.5$\times$0.5 & 0.013 \\
             & 0.4$\times$0.3 & 0.018 &              & 0.3$\times$0.3 & 0.013 \\
J0939+38     & 1.7$\times$1.4 & 0.041 & J1520+25     & 1.7$\times$1.3 & 0.031 \\
             & 0.7$\times$0.5 & 0.012 &              & 0.5$\times$0.5 & 0.013 \\
             & 0.4$\times$0.3 & 0.016 &              & 0.3$\times$0.3 & 0.013 \\
J0943+36     & 1.7$\times$1.4 & 0.015 & J1530+27     & 1.7$\times$1.3 & 0.076 \\
             & 0.7$\times$0.5 & 0.018 &              & 0.5$\times$0.5 & 0.020 \\
             & 0.4$\times$0.3 & 0.045 &              & 0.3$\times$0.3 & 0.024 \\
J1013+07     & 1.9$\times$1.4 & 0.043 & J1559+25     & 1.9$\times$1.3 & 0.061 \\
             & 0.8$\times$0.5 & 0.012 &              & 0.6$\times$0.5 & 0.023 \\
             & 0.5$\times$0.3 & 0.013 &              & 0.4$\times$0.3 & 0.023 \\ 
J1025+10     & 1.8$\times$1.4 & 0.044 & J1621+25     & 1.5$\times$1.4 & 0.036 \\
             & 0.7$\times$0.5 & 0.017 &              & 0.6$\times$0.5 & 0.013 \\
             & 0.4$\times$0.3 & 0.020 &              & 0.4$\times$0.3 & 0.011 \\        
J1040+09     & 1.8$\times$1.4 & 0.038 & J1628+25     & 1.5$\times$1.3 & 0.040 \\
             & 0.6$\times$0.5 & 0.017 &              & 0.6$\times$0.5 & 0.019 \\
             & 0.4$\times$0.3 & 0.012 &              & 0.4$\times$0.3 & 0.017 \\
J1136+51     & 1.5$\times$1.4 & 0.028 & J1658+25     & 1.5$\times$1.3 & 0.047 \\
             & 0.5$\times$0.5 & 0.012 &              & 0.5$\times$0.5 & 0.016 \\
             & 0.3$\times$0.3 & 0.011 &              & 0.3$\times$0.3 & 0.015 \\
J1213+50     & 1.4$\times$1.3 & 0.038 & J1703+24     & 1.6$\times$1.3 & 0.050 \\
             & 0.5$\times$0.5 & 0.027 &              & 0.5$\times$0.5 & 0.013 \\
             & 0.3$\times$0.3 & 0.060 &              & 0.3$\times$0.3 & 0.010 \\
  \hline
\end{tabular}
\label{tab2}

\smallskip
\small{Column description: (1) name; (2) beam size (arcseconds) at 1.4 GHz, 4.5 GHz and 7.5
  GHz in the three following rows; (3) rms (mJy beam$^{-1}$) at 1.4 GHz, 4.5 GHz
  and 7.5 GHz in the three following rows.}
\end{table}

\section{The sample and the VLA observations}
\label{sample}

Eighteen objects, whose main properties are listed in Table
\ref{tab1}, were randomly extracted from the FR0{\sl{CAT}} sample and
observed with the VLA. More specifically, we formed groups of
  three or four FR~0s located at small angular separations, as to
  reduce the telescope overheads and the time needed for observations
  of the calibrators. All groups were included in the schedule and
  five of them were actually executed. Because the observing strategy
  was only based on the location in the sky, this source selection does
  not introduce specific biases. The observed FR~0s reside in the redshift
range 0.019--0.050 and are well representative of the whole
FR0{\sl{CAT}} sample in terms of radio and AGN power, see
Figure~\ref{lrlo3}. This figure also shows the large deficit of radio
emission of FR~0s when compared to the FR~Is part of the 3C sample by
a factor ranging from $\sim$30 to $\sim$1000 in the same range of
bolometric AGN luminosity (represented by the [O~III] line
luminosity).

We obtained 8.67 hours of observations with the VLA in its A-array
configuration between December 27, 2016 and January 23, 2017. We observed the
18 objects in 5 separate scans, ranging from 1 to 2 hours, including 3 or 4
sources. Similar to the observation strategy for the VLA pilot study of 7
FR~0s used in \citet{baldi15}, the targets have been observed in L and C bands
splitting the exposure times in two.  While the L band configuration
corresponds to the default 1~GHz-wide band centred at 1.5 GHz, the C band was
modified based on our purposes. We divided the available 2-GHz bandwidth into
two sub-bands of 1 GHz centred at 4.5 and 7.5 GHz (hereafter C1 and C2 bands,
respectively). This strategy allows to obtain images in 3 different radio
frequencies in two integration scan. Each of the three bands was configured in
7 sub-bands of 64 channels of 1 MHz. Each source was observed for $\sim$10
minutes in both the L and C band, spaced out by the pointing to the phase
calibrators for 2-4 minutes. The flux calibrator was 3C~286 observed for
$\sim$6-7 minutes.

The data were calibrated by the CASA 5.0.0 pipeline v1.4.0, adding
further manual flagging to remove low-level radio-frequency
interferences and noisy scans to increase the general quality of the
data. The final imaging process was performed with {\it AIPS}
(Astronomical Image Processing System) package according to standard
procedures. The images were then produced in the L, C1, and C2 bands
from the calibrated data using the task {\it IMAGR}. The angular
resolutions reached in the three bands are, respectively, $\sim$ 1.7,
0.5, 0.3 arcsec. We self-calibrated the maps of the sources with flux
density $\gtrsim$5 mJy. The typical rms of the final images is $\sim$0.02 mJy,
measured in background regions near the target. We measured the flux
densities of the unresolved core components with the task {\it JMFIT}
and the total radio emission from the extended sources with the task
{\it TVSTAT}. In Tab.~\ref{tab2} we give the main parameters of the
resulting images. Fig.~\ref{maps} presents the maps of the extended
radio sources with radio contour level listed in Tab.~\ref{tab3}.

\begin{table}
  \caption{Contour levels (mJy beam$^{-1}$) for the sources presented in Fig. 2.}
\begin{tabular}{|l|c|l}
  \hline
  ID  & Freq. & Cont. Levels \\
  \hline
J0907+32   &  1.5 & 0.15$\times$(-1,1,2,4,8,16,32,64,128)   \\
           &  4.5 & 0.05$\times$(-1,1,2,4,8,16,32,64,128)   \\
           &  7.5 & 0.05$\times$(-1,1,2,4,8,16,32,64)   \\
J1213+50   &  1.5 & 0.36$\times$(-1,1,2,4,8,16,32,64,128,256)   \\
           &  4.5 & 0.20$\times$(-1,1,2,4,8,16,32,64,128,256)   \\
           &  7.5 & 0.27$\times$(-1,1,2,4,8,16,32,64,128,256)   \\
J1559+25   &  1.5 & 0.50$\times$(-1,1,2,4,8,16,32)   \\
           &  4.5 & 0.12$\times$(-1,1,2,4,8,16,32)   \\     
           &  7.5 & 0.12$\times$(-1,1,2,4,8,16,32)   \\   
J1703+24   &  1.5 & 0.135$\times$(-1,1,2,4,8,16,32,64)   \\
           &  4.5 & 0.035$\times$(-1,1,2,4,8,16,32,64)   \\
           &  7.5 & 0.03$\times$(-1,1,2,4,8,16,32,64)   \\
  \hline
\end{tabular}
\label{tab3}
\end{table}

\section{Results}
\label{results}

\begin{figure*}
\includegraphics[width=19cm,height=22cm]{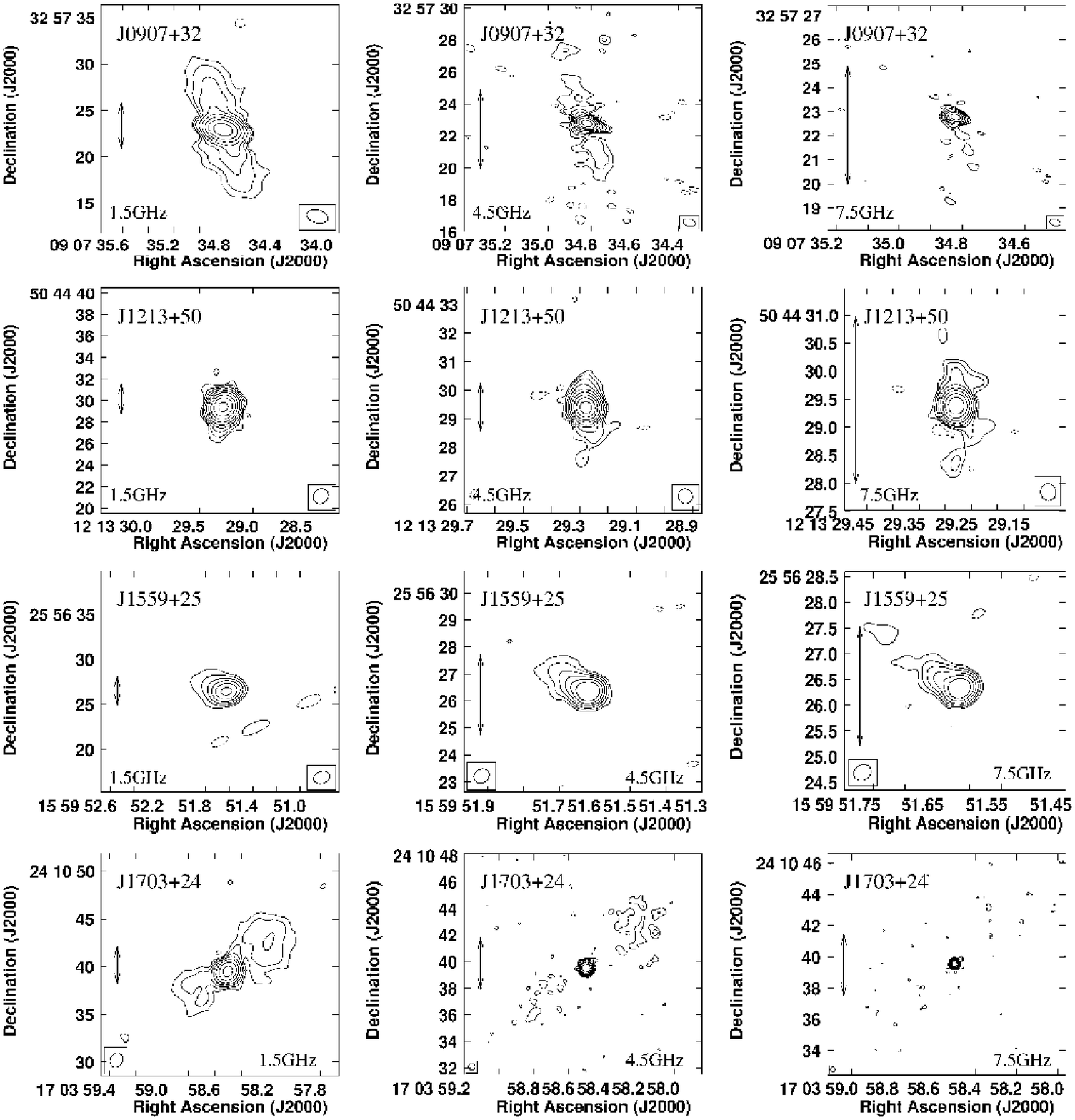}
\caption{[reduced-quality image for size limit] Radio images of the four FR~0 in which we detected extended
  emission. The three columns show the 1.5 GHz (left), 4.5 GHz
  (center), and 7.5 GHz (right), respectively. See Table~\ref{tab2}
  for angular resolutions and noise, and Table~\ref{tab3} for the list
  of contour levels. The arrowed lines represent a fixed scale bar
    for comparison for the three panels of each target.}
\label{maps}
\end{figure*}

The VLA observations show radio emission with flux densities in the
three bands ranging between 1 and 281 mJy (with typical errors of 0.04
mJy in L band and 0.02 mJy in the two C bands). Most of the sources
(14/18) appears unresolved down to a resolution of $\sim$0\farcs3
which corresponds to 100--300 pc.\footnote{In Appendix~\ref{notes}, we
  also show the radio maps at the three frequencies of IC~711,
  originally included in the FR0{\sl{CAT}} sample (and observed in
  this VLA project) but subsequently discarded based on the presence
  of extended emission in its NVSS image.} Four sources show instead
radio emission extended on a few arcseconds, on a scale of 2--14 kpc
(see Table \ref{tab4} and Fig. \ref{maps}). In J0907+32 two symmetric
jets reach a distance of $\sim$7$\arcsec$ ($\sim$14 kpc in largest
linear size, LLS). In J1213+50 the jets are best seen in the 4.5 GHz
image where they extend out to $\sim1\farcs5$ (LLS $\sim$2 kpc) from
the nucleus. Only one jet, $\sim3\arcsec$ long (LLS $\sim$3 kpc), is
detected in J1559+25. Finally, two diffuse radio structures are found
in J1703+24, with an angular size of $\sim$14$\arcsec$ (LLS $\sim$9
kpc). In the Table \ref{tab4} we also provide the total flux densities
on the entire radio source. For the four extended sources, we
estimated the jet counter-jet ratio measuring the brightness ratio in
two symmetric regions as near as possible to the nuclear emission by
avoiding the core and considering the surface brightness at similar
distance from the core. The brightness jet ratio range between $\sim$1
and 2, while one source is fully one-sided ($>$8).

Our high resolution VLA observations, obtained with a relatively
short exposure time, might miss faint extended emission. In
order to explore this possibility we compared the observed total VLA
1.5 GHz flux densities with those measured by FIRST and NVSS at larger
resolutions ($\sim$5 and $\sim$45 arcsec, respectively, see Table
\ref{tab4}). The flux densities measured in the L band from our maps are
generally consistent with those from the FIRST catalogue (derived from
observations between 1994 and 1999 for our sources), an indication
that we recovered most of the radio emission. Indeed, all but four
sources show differences of less than 20\% over a timescale of
$\sim$20 years. One of them (J1136+51) decreased in flux from 7.8 to
5.3 mJy. Three sources (namely J0943+36, 1025+10, and J1530+27)
instead increased their flux densities by a factor between 1.5 to 3.2.

All the 18 FR0s show a ratio between NVSS and FIRST flux densities very close
to unity ($0.89 < F_{\rm NVSS}/F_{\rm FIRST} < 1.11$) with only two
exceptions. One is J1559+25 with a ratio of $\sim$5: this is due to
the presence of a second compact source with a FIRST flux density of 117 mJy
located 28$\arcsec$ to the West, blended with our target in the NVSS
image. The second is J1703+24 with a ratio of 1.43: this is one of the
FR0s with extended emission, suggesting that some extended emission is
lost in the FIRST images.

\begin{figure*}
\includegraphics[scale=1.15]{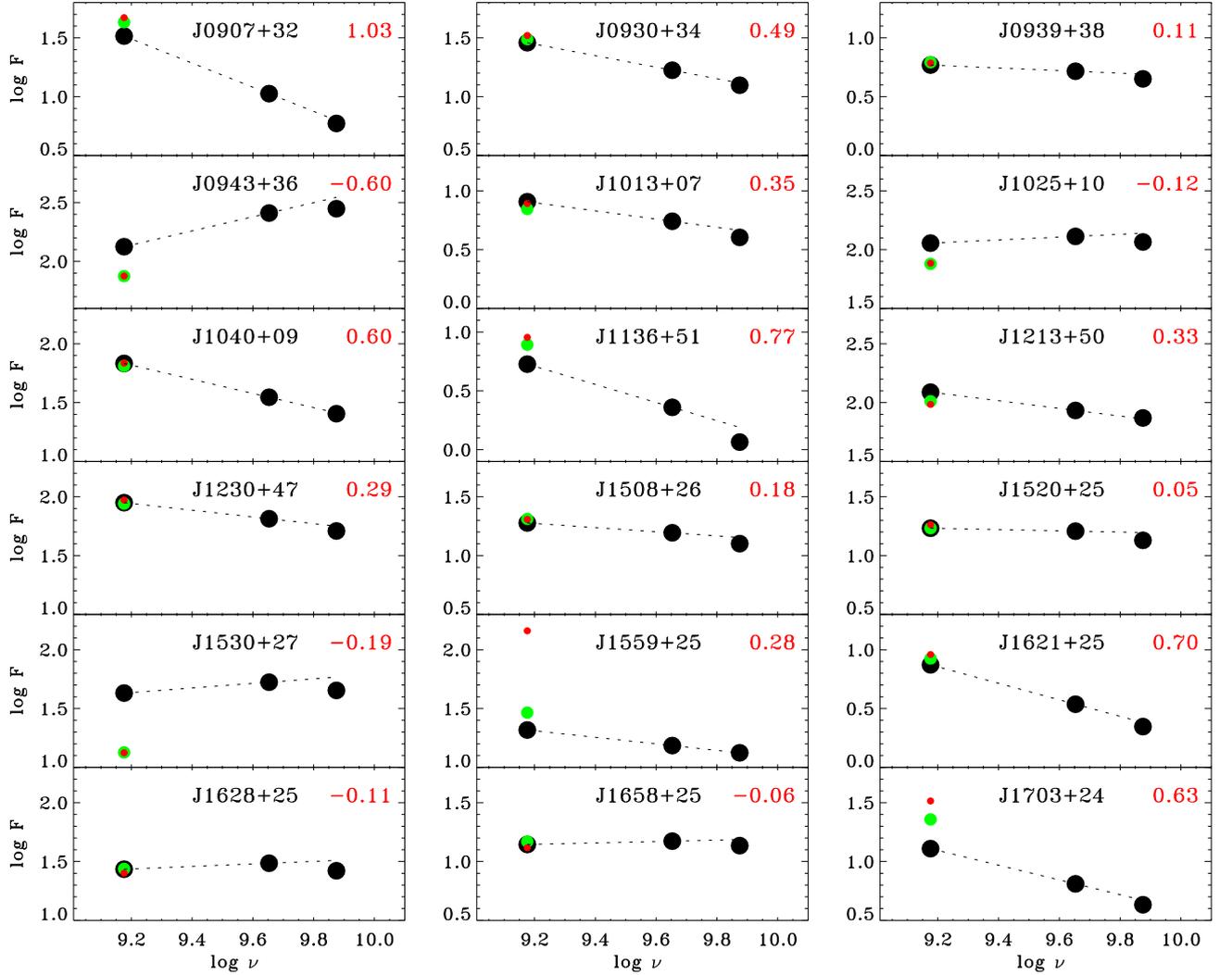}
\vspace{-1cm}
\caption{Radio spectra of the central components of the 18 FR~0s observed with
  the VLA. The black dots are the VLA data, while the red and green dots are the
  NVSS and FIRST flux densities, respectively. The dashed line represents the power
  law obtained from the 1.5 and 4.5 GHz VLA measurements; its slope is
  reported at the top right of each panel. Flux densities are in mJy, frequencies in
  Hz.}
\label{SED}
\end{figure*}

We also obtained matched-beam radio maps\footnote{We matched the
  C-band radio images at the resolution obtained in L band by using
  the parameter {\it uvtaper} in the {\it IMAGR} task in AIPS.} to
derive the radio spectra at the three frequencies for all sources. The
resolution-matched flux densities of the central components are
reported in Table \ref{tab4} and plotted in the radio spectra in
Fig.~\ref{SED}. The typical flux errors for the central components are
smaller than 0.1 mJy. Only for the brightest sources, exceeding
$\sim100$ mJy, they can be as high as 0.5 mJy.  We do not report them
in the Table for sake of clarity.

The spectral indices $\alpha$ between 1.5 and 4.5 GHz ($F_{\nu} \propto
\nu^{\alpha}$) cover a broad range, from $\sim$1 to -0.2: most spectra are
flat ($-0.2< \alpha < 0.4$ is measured for 11 objects), six are steep ($0.49 <
\alpha < 1.03$). In one case (J0943+36) the spectrum is strongly inverted,
with $\alpha$ = -0.6. If we consider the spectral indices measured between the
two adjacent bands, we typically observe a spectral steepening at 7.5 GHz,
with a median difference between the two indices of $\Delta\alpha \sim$
0.16. We also note that four flat-spectrum sources (1025+10, 1530+27 1628+25
1658+25) have slightly convex radio spectra: the flux density at 4.5 GHz is
typically $\sim$20 per cent higher than the one at 1.5 GHz and 15 per cent
higher than at 7.5 GHz.

\begin{table*}
\caption{Radio properties}
\begin{tabular}{|l|r|r|r|r|r|r|c|c|c}
  \hline
Name &NVSS &FIRST &F$_{1.5}$ &F$_{4.5}$ &F$_{7.5}$ &F$_{1.5, , \rm{tot.}}$  &Morph. &size &jet ratio\\
  \hline
J0907+32 & 46.9&  42.9&  32.7 & 10.61 &  5.92 & 42.5 &twin-jets &14 kpc &1.3$\pm$0.23\\
J0930+34 & 33.1&  30.8& 28.73 &  16.77& 12.51 &      &          &       &       \\
J0939+38 &  6.1&   6.2&  5.87 &   5.20& 4.48  &      &          &       &       \\
J0943+36 & 75.1&  74.9& 133.3 &  257.8& 280.4 &      &          &       &       \\
J1013+07 &  7.8&   7.0&  8.08 &   5.51& 4.02  &      &          &       &       \\
J1025+10 & 76.6&  75.7& 113.6 &  129.6& 116.2 &      &          &       &       \\
J1040+09 & 68.5&  64.7& 68.0  &  35.1 & 25.44 &      &          &       &       \\
J1136+51 &  9.0&   7.8&  5.33 &  2.29 & 1.16  &      &          &       &       \\
J1213+50 & 96.5& 102.7&  122.7& 85.8  & 74.0  &133.0 &two jets  & 2 kpc &2.0$\pm$0.7\\
J1230+47 & 93.8&  87.4& 89.0  &  65.0 & 51.2  &      &      &          &       \\
J1508+26 & 20.3&  20.5& 18.94 &  15.62& 12.63 &      &      &          &       \\
J1520+25 & 18.3&  17.1& 17.09 &  16.13& 13.48 &      &      &          &       \\
J1530+27 & 13.3&  13.4& 42.8  &  52.9 & 45.1  &      &      &          &       \\
J1559+25 &155.9&  29.2&  20.78& 15.31 & 13.30 & 24.5 & one jet  & 3 kpc &$>$ 8\\
J1621+25 &  9.1&  8.4 &  7.44 &  3.44 & 2.22  &      &      &          &       \\
J1628+25 & 25.2& 27.4 & 27.22 &  30.59& 26.40 &      &      &          &       \\
J1658+25 & 13.1& 14.8 & 13.97 &  14.91& 13.65 &      &      &          &       \\
J1703+24 & 32.7& 22.8 &  12.87& 6.46  &  4.29 & 17.7 & two lobes& 9 kpc &1.0$\pm$0.4\\
  \hline
\end{tabular}
\label{tab4}

\medskip
\small{Column description: (1) source name; (2) and (3) NVSS and FIRST
  1.4 GHz flux density (mJy); (4), (5), and (6) fluxes of the nuclear
  components from matched-beam images at the three observing
  frequencies (mJy), (4) total flux at 1.5 GHz flux density (mJy); (8)
  morphology for the extended sources, (9) largest linear size, and
  (10) brightness ratio between the opposite structures for the
  extended sources}.
\end{table*}

\begin{figure}
\includegraphics[scale=0.5]{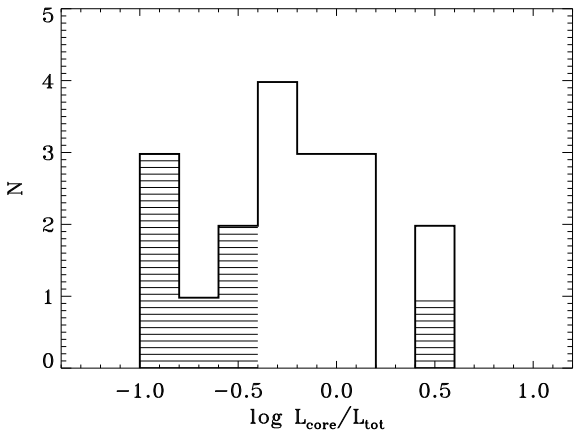}
\caption{Distribution of core dominance $R$, i.e., the ratio between
  the flux density of the central component at 7.5 GHz and the NVSS flux
  density at
  1.4 GHz. The hatched histogram corresponds to upper limits on $R$ for
  the sources with no radio core detection.}
\label{coredom}
\end{figure}

\begin{figure}
\includegraphics[scale=0.55]{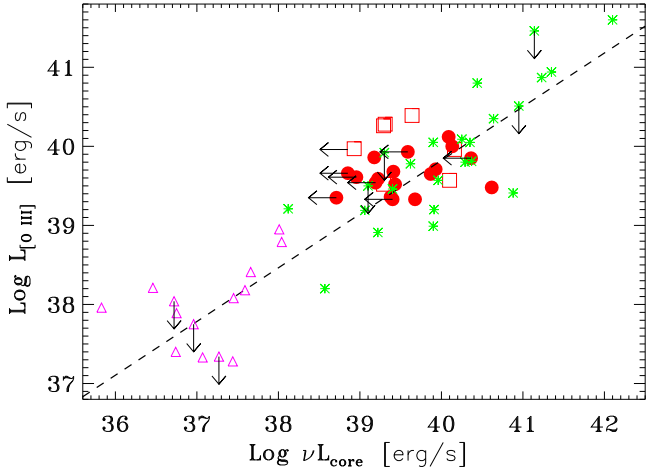}
\caption{Core radio power vs. [O~III] line luminosity (erg s$^{-1}$)
  for CoreG (pink triangles), 3CR/FR~I radio-galaxies (green
  asterisks), and FR0s (red dots) from this study and the 7 FR~0s (red
  squares) from the VLA pilot study \citep{baldi15}. The line
  indicates the best linear correlation found for 3CR/FR~Is.}
\label{coreo3}
\end{figure}

Interestingly, results of VLBI observations for four of FR~0s studied here
have been recently presented by \citet{cheng18}. J0943+36, the
inverted-spectrum source, shows an unresolved VLBI core with an inverted
spectrum and significant variability, with a flux density increasing from 0.17
to 0.25 Jy over four years. J1213+50, a twin-sided jet FR~0,
show a similar morphology but on mas-scale and perpendicular to the
VLA jets. J1230+47, a flat-spectrum FR~0, shows twin jets on
mas scale with the VLBI. J1559+25, the one-sided FR~0, appears
similarly one-sided on mas scale with the VLBI with no sign of variability
($<20$\%).

One of the aims of this study is the measurement of the radio core. In
the four extended sources a compact central component is always
clearly visible. However, the spectrum of the central source of, e.g.,
J0907+32 is very steep ($\alpha\sim1$) indicating a substantial
contribution from optically thin, extended emission. We must then rely
on the radio spectra to isolate the core emission. We considered as
core-dominated the 11 sources with central component characterised by
a flat spectrum ($\alpha\lesssim0.4$). For the remaining objects the
7.5 GHz flux density is adopted as upper limit to the core. In \citet{baldi15}
we noticed several sources with a flattening of the radio spectrum
between 4.5 and 7.5 GHz with respect to that measured between 1.5 and
4.5 GHz and interpreted this behaviour as the emerging of a flat core
at the higher frequencies. In the observations we are presenting here,
there is not a clear evidence of a flattening of the radio spectra at
7.5 GHz: the extrapolation of the 1.5 $-$ 4.5 GHz agrees or slightly
over-predicts the actual measurement at 7.5 GHz by $\sim$20 per cent.

The distribution of core dominance of our sources, $R$, defined as the ratio
between the nuclear emission at 7.5 GHz and the total NVSS flux density at 1.4
GHz\footnote{Except for J1559+25 for which we used the FIRST data because of
  the contamination to the NVSS flux density from the nearby source.} is presented in
Fig. \ref{coredom}. The 11 sources with a flat spectrum have $-0.4 \lesssim
{\rm log} R \lesssim 0.5$. The upper limits on $ {\rm log} R$ for those with a
steep spectrum are in the range -0.9 to -0.4.  This distribution is
significantly different from from that of 3CR/FRIs (with a $>$99.9\%
probability, according to a Kolmogorov Smirnoff test) while it is not
distinguishable from the $R$ distribution of CoreG and the other FR~0s studied
in the previous VLA project \citep{baldi10a,baldi15}.\footnote{We estimated
  $R$ as the ratio between the 7.5 GHz core emission and the total 1.4 GHz
  flux density, while for the 3CR/FRIs sources we used the 5 GHz core flux
  density against the
  1.4 total flux density. However, since the radio core emission has generally a flat
  spectrum this quantity is only weakly dependent on the frequency used for
  the core measurement and this comparison is robust.}

We can now include our sources in the $L_{core} (= L_{\rm 7.5 GHz})$
vs. \loiii\ plane similarly to what done in \citet{baldi15} for the
VLA FR~0 pilot study. This diagnostic plot compares the radio core
energetics with the line luminosity, adopted as a proxy for the AGN
luminosity \citep{heckman04}. The FR~0s lie in the same region
populated by the lower luminosity 3CR/FR~Is and follow the same
core-line relation found for FR~Is and CoreG. This relation further
strengthens the similarity between the nuclei of FR~0s and
FR~Is. However, we remind that for seven of them we only derived upper
limits to their radio core flux densities for the lack of a clear
flat-spectrum unresolved core. Although the point scatter of FR~0s is
consistent with that of FR~Is, the presence of upper-limits on the
core measurements in the correlation challenges the FR~0--FR~I
similarity scenario.

\section{Discussion}
\label{discussion}
 
Despite the fact that FR~0s share the same properties with FR~I radio
galaxies from the nuclear and host point of view, FR~0s differ from
the other FR classes for their remarkable lack of significant extended
radio emission, and for being dominated by a sub-kpc scale
flat-spectrum component. In \citet{baldi15} and \citet{baldi18a} we
discussed various scenarios to interpret this unique feature of FR~0
and we now review them in the light of the results obtained with the
new VLA observations.
  
\begin{figure*}
\includegraphics[scale=0.99]{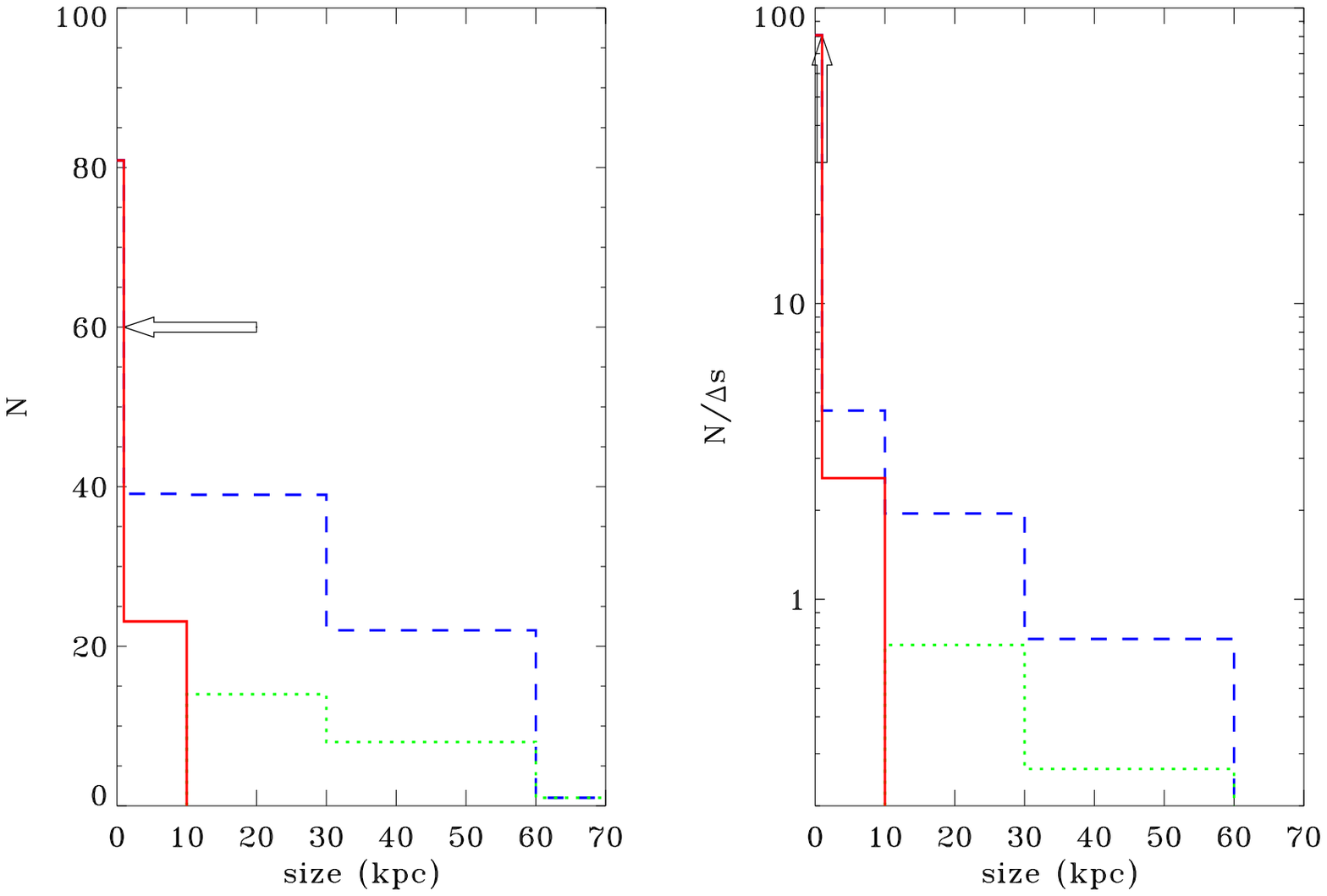}
\caption{Left: size distribution of SDSS/NVSS LEG radio galaxies with $z<$0.05
  (dashed blue histogram). We separate the contribution of FR~Is (dotted
  green) and FR0s (solid red): the first bin contains the 14 unresolved FR0s,
  with a size limit of $<$1 kpc (scaled by our VLA coverage of the
  FR0{\sl{CAT}} sample, i.e., by 104/14) the second bin represents the FR0s
  found to be extended from the VLA observations. Right: size distribution
  normalised by the bin size in kpc (note the logarithmic scale of the
  vertical axis). The first bin is the lower limit corresponding to the
  unresolved FR~0s.}
\label{size}
\end{figure*}

\citet{baldi18a} conclude that a scenario in which all FR~0s are young
radio sources that will eventually evolve into extended radio source
does not reconcile with the relative number densities of these
classes. Thanks to the new high-resolution observations, we can
further test this possibility, by exploring the distribution of radio
sizes of FR~0s. In Fig.~\ref{size} we show the size distribution of
all 182 SDSS/NVSS LEG radio galaxies with $z<$0.05, 104 of which are
included in FR0{\sl{CAT}}. The fraction of FR0s unresolved in the VLA
observations is $\sim78$\%, for which we set a conservative size limit
of 1 kpc (represented by the arrow in the Fig.~\ref{size} left panel),
and they represent $\sim44$\% of the whole population of radio
emitting AGN within this volume. The remaining sources extend up to
$\sim$100 kpc and about one third of them are FR~Is, part of the
FRI{\sl{CAT}} and sFRI{\sl{CAT}} samples. In the right panel of this
Figure we present the number of objects in each bin, divided by the
bin size in kpc. As already discussed in \citet{baldi18a}, in case of
constant expansion speed all bins should be equally
populated. Conversely, the number density of sources with size
$\lesssim$1 kpc exceed by $\sim$2 orders of magnitude that of the
extended sources indicating that FR~0s do not generally grow to become
large radio galaxies.

The distribution of sizes also sets strong constraints of the
interpretation that FR~0 are short-lived recurrent sources. As we
mentioned in \citet{baldi15}, the results obtained by \citet{shin12}
indicate that the more massive galaxies spend a larger fraction of
their time in active states than satellite galaxies. The possibility
that this is the origin of the strong peak in the size distribution
appears to rather contrived considering the relatively small
differences in host galaxy masses between FR~0s and FR~Is, on average
only a factor 1.6 and with substantial overlap between the two classes
\citep{baldi18a,miraghaei17}. Nevertheless, it is correct to mention
that a low/moderate amplitude radio flux density variation has been
detected for a few FR~0s over time scale of years
\citep{cheng18}. This variability is not necessary associated with a
nuclear recurrence, but a phenomenon expected within an evolutionary
scenario of the source \citep{morganti17}.

Another scenario for the origin of FR~0s is related to the jet launching
region. The similar nuclear luminosities of FR~0s and FR~Is and the FR~0 radio
morphologies we observed suggest that within the radio core, less than 1 kpc,
the jets of FR~0s should start relativistic \citep{cheng18}. However, at
larger scales ($>$ 1 kpc), we envisaged that the jet $\Gamma$ factors of FR~0s
are lower than in FR~Is by affecting their ability to the penetrate the host's
interstellar medium and transforming the relativistic jets to turbulent flow
not far outside the optical core of the galaxy as seen in nearby FR~I radio
galaxies (e.g. \citealt{killeen86,venturi93,bicknell95}). Nevertheless, no
evidence of deceleration sites along the jet on different scales has been
observed in our FR~0 sources. We can test the possibility of a jet
deceleration and the bulk jet speed by measuring the sidedness of jets in
FR~0s and to compare it with that of FR~Is. By considering together this new
dataset and the results of the pilot program there are now five FR~0s with
rather symmetrical extended or slightly sided structures (jet brightness ratio
1--2, J0907+32, J1213+50, J1703+24 in this work and ID~547 and ID~590 from
\citealt{baldi15}).  We also found one highly asymmetric source but the
minimum jet Lorentz factor to obtain the observed flux ratio between the jet
and counter-jet of $\gtrsim 8$ is only $\Gamma \gtrsim 1.1$, not necessarily
indicative of a highly relativistic jet.  From the comparison between our VLA
and VLBI images from \citet{cheng18} for 5 FR~0s, we learnt that, when
observed, sub-kpc scale jets are typically present also at mas scale. However,
for our source (J1213+504) the mas-scale jets appear perpendicular to those
observed with VLA, possibly due to jet procession, while in one compact FR~0
(J1230+47), two pc-scale twin jets emerge, pointing to a sub-relativistic jet
speed on sizes smaller than what probed by the VLA. An inferred mildly
relativistic jet bulk speed on sub-kpc scale is in agreement with the Doppler
boosting factors estimated by \citet{cheng18} from VLBI observations. The
modelling of the multiband spectrum of the first FR~0 detected in $\gamma$-ray
with Fermi \citep{grandi16} with standard beamed and misaligned jet models
\citep{maraschi03,ghisellini05} produce bulk jet $\Gamma$ factor 2 -- 10
\citep{tavecchio18}. Clearly, the statistic on the jets asymmetries in FR~0s
is still insufficient to draw a firm conclusion on their speed, but an
evidence of mildly-relativistic jet speed is increasing.

Another tool that can be used to measure the jets speed is to consider
the dispersion of the core powers with respect to a quantity
independent of orientation, such as the emission line luminosity, see
Fig. \ref{coreo3}. Because there might be a substantial level of
intrinsic scatter, this approach can provide us with an {\sl upper
  limit} on the jet $\Gamma$. However, due to the presence of a
significant fraction of non detections of flat cores, we can only
derive a {\sl lower limit} to the cores dispersion. Higher frequency
observations are needed to isolate the core emission also in the
sources with steep spectra and fully exploit this method.

We must also mention that a different matter content of the FR~0 jets from the
other FR classes might account for the reduced jet extension observed in
FR~0s.  Lighter jets, possibly composed mainly of electron-positron pairs
\citep{ghisellini12} would correspond a much lower jet power, again, hampering
the formation of large scale radio structures, with respect to what is
predicted by a leptonic model \citep{maraschi03}.

While the majority of FR~0s conform with the idea that they are compact flat
spectrum sources, some of them show steep spectra. Interestingly, the
  fraction of flat and steep sources in our sample (with a two to one
  ratio) is the same found by \citet{sadler14} in their sample of FR~0 LEG,
  despite the rather different selection criteria. FR~0s certainly are a
mixed population low-power radio sources, one expected contaminant being a
small population of genuinely young FR~Is, as also suggested by
\citet{cheng18}. 

We must mention that four sources show a slightly convex
radio spectra and one source (J0943+36) has an inverted spectrum, similar to
what seen in GHz peaked sources (GPS, \citealt{peacock82}). Nevertheless, VLBI
observations \citep{cheng18} did not resolve a mas-scale double-lobe structure
as expected by this class of radio sources. Therefore, its particular radio
behaviour requires more attention to investigate whether its compactness and
spectrum reconciles with being a young radio galaxy, which expands at a very
low rate due to its much lower power with respect to GPS \citep{odea98} or,
instead, is consistent with what expected from young FR~Is.

\section{Summary and Conclusions}
\label{conclusion}

We presented new VLA observations at array A at three frequencies
1.5, 4.5 and 7.5 GHz for a sample of 18 FR~0 radio galaxies selected
from the FR0{\sl{CAT}} sample. At the highest angular resolution
($\sim$0.3 arcsec), most of the sources are still unresolved with
with $\sim$80 per cent of the total radio emission enclosed in the
core (i.e., they are more core-dominated than FR~Is by a factor
$\sim$30).  Only four objects show extended emission up to 14 kpc,
with various morphologies: one- and two-sided jets, and double
lobes. Six have steep radio spectra, one shows an inverted spectrum,
while 11 are flat cores. Four of the flat-spectrum cores in FR~0s are
actually slightly convex, with a small steepening between 4.5 and 7.5
GHz with respect to the slope measured at lower frequencies.

For 11 the sources, where a radio core is detected, the core and
emission line luminosities correlate following the relation valid for
FR~I and CoreG radio galaxies, respectively at high and low
luminosities. However, for 7 sources due to their their steep spectra
and morphology, we can give only an upper limit to the radio core
luminosity, which is still in agreement with the correlation. More
high-resolution data are necessary to isolate their core
components. Adding the 7 FR~0s studied in a previous VLA exploratory
pilot \citet{baldi15}, we confirm the similarity between the radio
cores of FR~0s and FR~Is and the absence of strong beaming effect in
the FR~0 cores.

The size distribution of nearby $z<0.05$ RL AGN shows, thanks to the
new observations, a strong peak for sources smaller than $\sim
1$kpc. This rules out the possibility that FR~0s are young sources
that will all evolve into more extended radio galaxies and it also
disfavours the interpretation that they are recurrent sources, with
very short periods of activity. The most likely possibility is that
FR~0s are associated with jets that are mildly relativistic at sub-kpc
scale.

The combination of high-resolution, large collecting area, and wide
frequency range of the new generation of radio telescopes (e.g., SKA,
ngVLA, LOFAR) will enable to satisfy the increasing interest in
low-power radio sources \citep{nyland18}, and, consequently, in FR~0s
\citep{whittam17}. The advent of larger radio surveys will give the
opportunity to study the pc-scale emission of FR~0s and test whether
their jets have smaller bulk jet speed than the other FR classes on
firmer statistical grounds, by expanding the size of the observed
sample with high quality radio observations.  Higher frequency
observations are also needed to isolate the core emission in sources
in which the radio spectra is steep and detect their jets. On the
opposite side, lower frequency observations will be very useful to
establish whether among them there are other inverted-spectrum
sources, which will provide indications on the nature of the FR0
population: slow jets or young radio sources?

\section*{Acknowledgments}
The National Radio Astronomy Observatory is a facility of the National
Science Foundation operated under cooperative agreement by Associated
Universities, Inc. RDB acknowledge the support of STFC
under grant [ST/M001326/1].  We thank Francesca Panessa, Graziano
Chiaro and Matt Malkan for a helpful discussion on the nature of
compact radio sources. We also thank the reviewer for the useful
comments, which helped us improve the quality of the manuscript.

\bibliography{my}

\appendix

\begin{figure*}
\centering
\includegraphics[width=19cm,height=5cm]{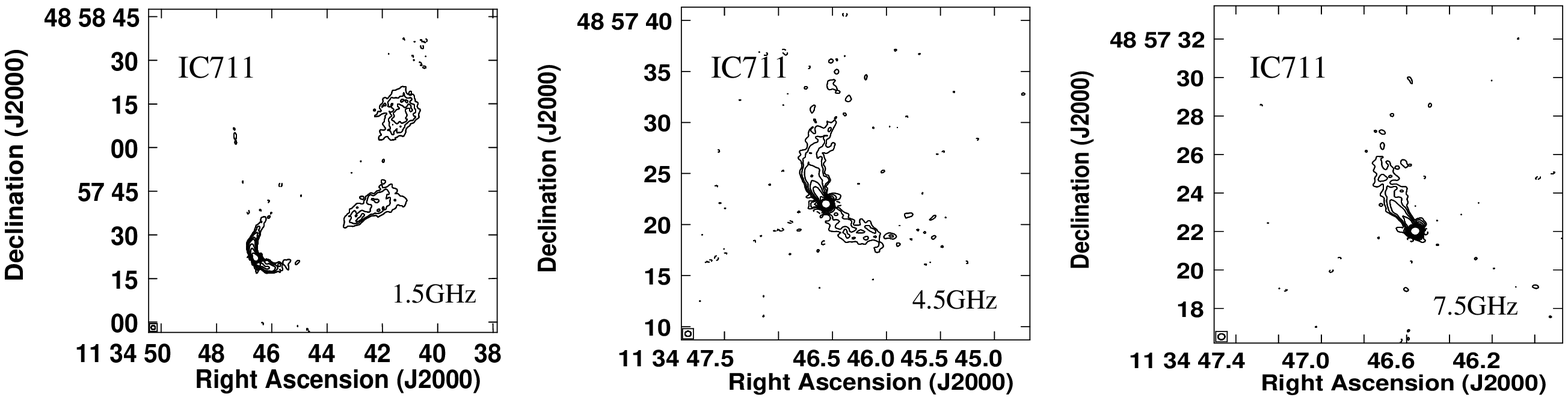} 
\caption{VLA images of SDSS~J113446.55+485721.9 (aka IC~711) at 1.5 (top),
  4.5 and 7.5 GHz (bottom left and right, respectively. Contour levels are
  drawn at: 0.17$\times$(-1,1,1.5,2,3,4,8,16,32)  mJy beam$^{-1}$ at 1.5 GHz; 0.04$\times$(-1,1,2,4,8,16,32,64,128)  mJy beam$^{-1}$ at 4.5 GHz; 0.05$\times$(-1,1,2,4,8,16,32,64,128)  mJy beam$^{-1}$ at 7.5 GHz.}
\label{ic711} 
\end{figure*}

\begin{figure*}
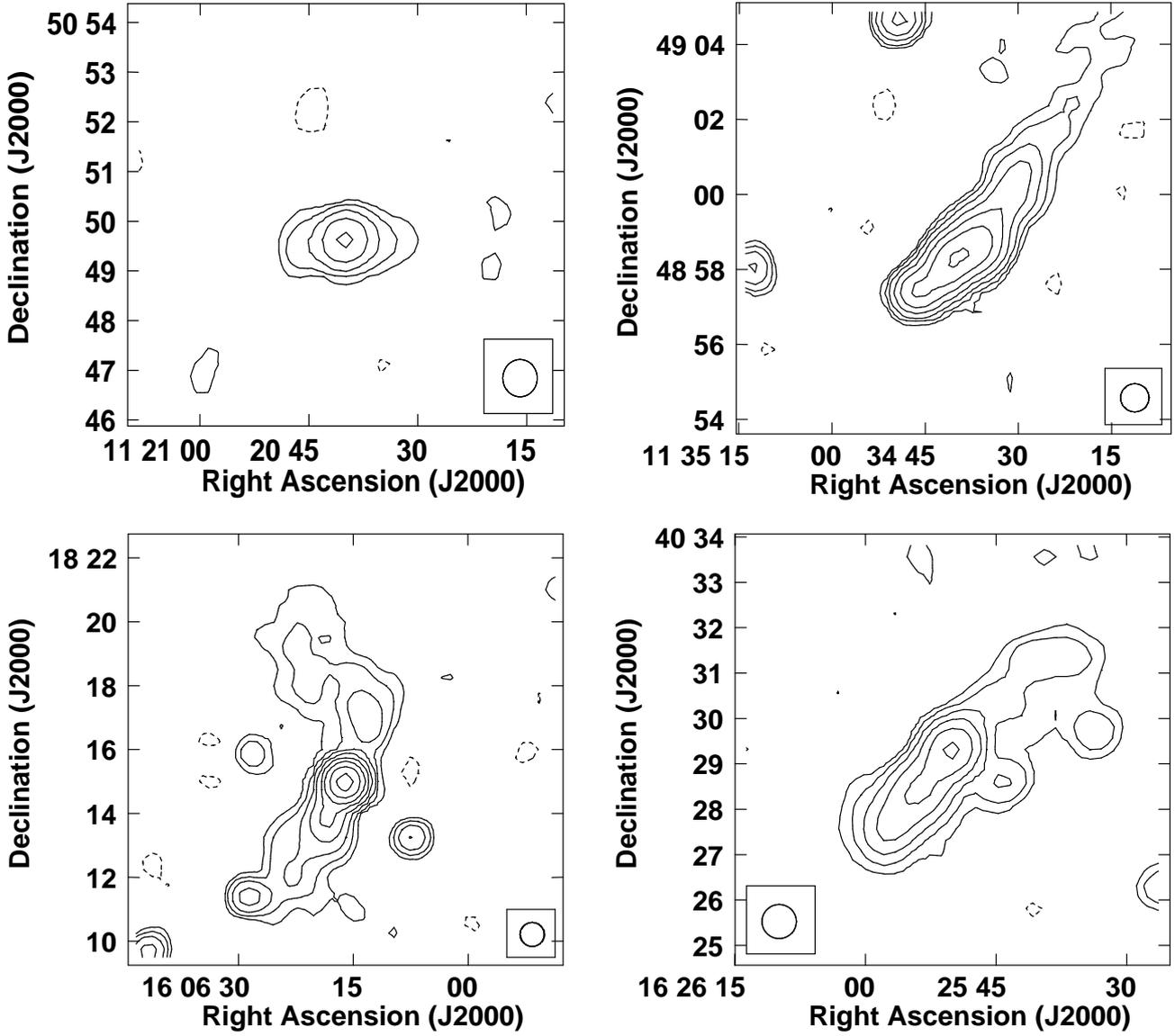

\includegraphics[width=8.8cm,height=7.8cm]{nvssJ1120+50bis.ps}
\includegraphics[scale=0.46]{nvssJ1134+48.ps}
\includegraphics[scale=0.46]{nvssJ1606+18.ps}
\includegraphics[scale=0.46]{nvssJ1625+40.ps}
\caption{NVSS images of the four sources erroneously included in the
  FR0{\sl{CAT}} but clearly showing extended radio emission and now
  excluded from the catalogue: from top left to bottom right SDSS~J1120+50 SDSS~J1134+48 aka IC~711, SDSS~J1606+18, and SDSS~J1625+40). The contour levels of the maps are (-1,1,2,4,8,..) mJy beam$^{-1}$.}
\label{NVSS}
\end{figure*}

\section{The case of IC~711 and an update of the FR0{\sl{CAT}} }
\label{notes}

Most of the 108 sources in the FR0{\sl{CAT}} catalogue show a ratio
between the FIRST and NVSS flux densities between 0.8-1.2, indicating
that, in general, in FR~0s there is no significant amount of extended
low-brightness radio emission lost due to the missing short baselines
of the FIRST observations. However, individual exceptions might
exist. In Fig.~\ref{ic711} we present the images obtained as
part of our program of VLA observations of SDSS~J113446.55+485721.9
which shows a pair of bent asymmetric jets extending for
$\sim$90\arcsec\ from the radio core. This source, also known as
IC~711, is actually the longest head-tail radio source known
\citep{vallee76}, extending over $\sim$ 500 kpc \citep{vallee87}. Its
extended nature is clearly seen also in the NVSS image (see
Fig. \ref{NVSS}).

The NVSS images were not used for the FR~0s selection because of their
low spatial resolution (45$\arcsec$) and the resulting high level of
confusion. Nonetheless, the case of IC~711 indicates that they can
still be used to improve the sources morphological classification. We
then retrieved the NVSS images of all 108 FR~0s in FR0{\sl{CAT}}
looking for extended emission. Elongations in the iso-contours are
rather common, usually extending just in one direction. As expected,
in most cases this is the result of confusion, due to the presence of
a nearby radio source, well visible in the FIRST images. Nonetheless,
there are four sources in which the radio emission is clearly extended
. Their NVSS images are shown in Fig. \ref{NVSS}. These sources
(namely SDSS~J112039.95+504938.2 SDSS~J113446.55+485721.9 aka IC~711,
SDSS~J160616.02+181459.8, and SDSS~J162549.96+402919.4) are not
genuine compact FR~0s and should be then removed from the
FR0{\sl{CAT}}catalogue, which now counts 104 sources.

Regarding IC~711, the core flux densities measured from the VLA
observations are 25.92 $\pm$ 0.03 at 1.5 GHz, 24.30 $\pm$ 0.02 at 4.5
GHz, and 23.23 $\pm$ 0.01 at 7.5 GHz, respectively, showing a flat
spectrum with $\alpha$ = - 0.07.  In this source the jets are
  highly asymmetric in the innermost regions, but their brightness
  ratio decreases at larger distances, possibly due to a deceleration
  from relativistic to subsonic \citep{bicknell95}.

\label{lastpage}
\end{document}